\documentclass{aa} 

\usepackage{graphicx}
\usepackage{txfonts}
\usepackage[
colorlinks=true,
citecolor=blue,
]{hyperref}
\begin{document}

   \title{Ring dynamics around an oblate body with an inclined satellite: The case of Haumea}

   \subtitle{}

   \author{
           Francesco Marzari 
          }

   \institute{Department of Physics and Astronomy, 
              University of Padova, via Marzolo 8, I-35131, 
              Padova, Italy\\
              \email{francesco.marzari@pd.infn.it}
             }

   \date{Received ....; accepted .....}

 
  \abstract
  {The recent discovery of rings and massive satellites around minor bodies and dwarf planets suggests that they may often coexist, as for example around Haumea.}
   {A ring perturbed by an oblate central body and by an inclined satellite may  disperse on a short timescale. The conditions under which a ring may survive are explored both analytically and numerically.  }
   {The trajectories of ring particles are integrated under the influence of the gravitational field of a triaxial ellipsoid and (a) massive satellite(s), including the effects of collisions. }
   {A ring initially formed in the equatorial plane of the central body will be disrupted if the satellite has an inclination in the Kozai--Lidov regime ($ 39.2^o < i < 144.8$). For lower inclinations, the ring may relax to the satellite orbital plane thanks to an intense collisional damping. On the other hand, a significant J2 term easily suppresses the perturbations of an inclined satellite within a critical semi--major axis, even in the case of Kozai--Lidov cycles. However, if the ring is initially inclined with respect to the equatorial plane, the same J2 perturbations are not a protective factor but 
instead disrupt the ring on a short timescale. The ring found around Haumea is stable despite the rise in the impact velocities that is due to the asymmetric shape of the the body and the presence of a 3:1 resonance with the rotation of the central body. }
   {A ring close to an oblate central body should be searched for in the proximity of the equatorial plane, where the J2 perturbations protect it against the perturbations of an external inclined satellites. In an inclined configuration, the J2 term is itself disruptive. }


   \maketitle
%

\section{Introduction}

 In the Kuiper Belt, all six of the largest bodies with diameters greater than 1000 km have a satellite system with a secondary--to--primary mass ratio ranging from $10^{-1}$ to $10^{-3}$.  According to \cite{arakawa2019}, these satellites likely formed via giant impacts in the early stage of solar system formation.  Another interesting feature, possibly related to the outcome of a less energetic collision, is the possible presence of a ring around some of these bodies. The two centaurs 10199 Chariklo and 2060 Chiron   are known to have ring systems detected by stellar occultation \citep{ribas2014,ortiz2015,ruprecht2015}; recently, \cite{ortiz2017} found a ring around the trans-Neptunian dwarf planet Haumea, which is also orbited by two satellites, Namaka and Hi'iaka. These rings may have originated from the ejection of cratering fragments from the parent body, the disruption of a satellite \citep{pan2016}, or the partial tidal disruption of the mantle during a close encounter with a giant planet \citep{ryuki2016}. 
 
 The long--term survival of a ring around a body possessing one or more inclined satellites can be jeopardized by their secular perturbations, which would act on a much shorter timescale than the viscous spreading \citep{salmon}. These perturbations would be particularly disruptive for the ring coherence in the presence of Kozai--Lidov configurations, which force wide oscillations in both eccentricity and inclination \citep{naoz2016}. However, even a mild inclination may lead to a misalignment of the ring particle orbits once the node longitudes are randomized due to the difference in semi--major axes.

There are two ways a ring can survive in the presence of a massive satellite on an inclined orbit. In the absence of strong Kozai--Lidov perturbations, the collisional damping may be intense enough to align the ring to the satellite orbital plane, where it will survive without losing coherence.  Alternatively, even for high inclinations of the satellite,  the J2 term can suppress the satellite perturbations, keeping the ring flat in the equatorial plane, if the central body is oblate and the ring lies in its equatorial plane.

I will explore these scenarios and, in particular, I will focus on the contribution from the central body oblateness to the stability of the ring.  The dynamics of the ring around the dwarf planet Haumea, a rapidly rotating triaxial ellipsoid with an elongated shape, will also be investigated. I will focus on the interplay between different evolutionary mechanisms that act on the ring, such as collisions and the gravitational perturbations from the irregular shape of the central body and from its two satellites. 

In Sect. 2 I describe in detail the numerical model exploited to perform the numerical simulations of the ring evolution. Section 3 is devoted to the rivalry between the Kozai--Lidov cycles and the J2 perturbations by the central body in the presence of a highly inclined satellite. The less dramatic but still potentially harmful effects of secular perturbations from a mildly inclined satellite are explored in Sect. 4. In Sect. 5 I illustrate the scenario where the ring is not initially placed in the equatorial plane of an oblate central body and the  J2 perturbations lead to a quick disruption of the ring.  Section 6 focuses on the dynamics of Haumea's ring, while the results of this whole dynamical exploration are discussed and commented on in Sect. 7.  
 
 \section{The numerical model}
 
The ring evolution was modeled by numerically integrating the trajectories of the ring particles, treated as massless bodies, that evolve in the gravity field of a central body of mass $M$ and one or more external perturbers (satellites). The 15th order Radau integrator \citep{radau1985} was used to better deal with the fast changes in the gravitational force acting on the particles. The central body was assumed to be a triaxial ellipsoid with three independent semi--axes ,$r_a$, $r_b$, and $r_c$. The potential of this body was computed with  
 MacCullagh’s formula \citep{murray-dermottSS}: 
 
 \begin{equation}
 V = - \frac{G M}{r} - \frac{G(A+ B+C -3I)} {2 r^3}
, \end{equation}
\label{Maccullagh1}

where 

 \begin{equation}
 I  = \frac {(A x^2 + B y^2 +Cz^2)} {r^2}
.\end{equation}
\label{Maccullagh2}  

The inertial moments $A,B,$ and $C$, in the case of a triaxial ellipsoid, are related to the semi--axes by the following relations:

\begin{equation}\begin{split}
A & = \frac{4}{15} \pi \rho abc (r_b^2 + r_c^2) \\
B & = \frac{4}{15}\pi \rho abc (r_a^2 + r_c^2) \\
C & = \frac{4}{15}\pi \rho abc (r_a^2 + r_b^2)
\end{split},\end{equation}
\label{Maccullagh3}

where $\rho$ is the body bulk density. In this case, $J_2$ can be obtained as \citep{turcotte}

\begin{equation}
 C - A = J_2 M r_a^2
.\end{equation}
\label{j2_ellipsoid}

In the code, all possible two--body encounters were tested at each time step. Since the number of particles is limited by the CPU load and the number of impacts would be too low if I took the particles’ true sizes, an “inflated radius” was adopted  for each particle \citep{brahic1976}. This inflated radius was made large enough to provide reliable impact statistics.  The evolutionary speed was the same if the number of bodies was scaled by the ratio of the real 
particle radius $r$ and the inflated radius $R$ according to $
N (R/r)^2$.

Each collision was treated as an inelastic rebound modeled with the algorithm of \cite{brahic1977}, which is based on the assumption that the 
self--attraction of the particles can be neglected and that all particles have the same radius and mass. During a collision, the radial relative velocity was damped with a rebound coefficient $\eta < 1$ while the grazing one was left unaltered (frictionless spheres). The sizes of the ring particles were scaled as suggested by \citet{brahic1976}, \citet{theb1998}, and \citet{charnoz2001}. 

\section{Kozai--Lidov cycles suppression by the J2 perturbations}

If the body has a massive satellite on a highly inclined orbit, the Kozai--Lidov perturbations may be a relevant disruption mechanism for a ring in the equatorial plane of the central body.  In this scenario, the Kozai--Lidov perturbations on the ring particles lead to a fast disintegration of the ring coherence  if the central body is spherical, and most particles are ejected or impact the central body. This occurs even if the mutual collisions between the ring particles tend to damp the eccentricity and inclination. 

I first performed a test numerical simulation with  a spherical central body ($r_a=r_b=r_c$) that has the same mass as the TNO (trans-Neptunian object)
Haumea ($M_c = 4.006 \times 10^{21}$ kg) and a satellite similar to Hi'iaka with a mass equal to $m_s = M_c/223 $. The orbit of the satellite has an eccentricity of 0.05 and an inclination of $55^o$.  A set of 10000 particles simulating the ring were initially distributed in the equatorial plane of the central body between 5000 and 10000 km, with  an eccentricity lower than $10^{-3}$ and an inclination in the range of $0^o$ to $1^o$. An inflated diameter of 10 km was adopted, translating to  approximately $10^9$ decimeter--sized particles according to the 
$
N  (R/r)^2$ scaling. The rebound coefficient $\eta$ was set to 0.3. 

After 10000 days of evolution, and despite the damping effects of the collisions, all the ring particles end up either on hyperbolic orbits or impact the central body because of the wide Kozai--Lidov oscillations of eccentricity and inclination. In this scenario, the ring is disrupted and could not survive longer than a few thousand days.  

\begin{figure}
\hskip -1.9 truecm
\includegraphics[width=1.2\columnwidth,angle=-90]{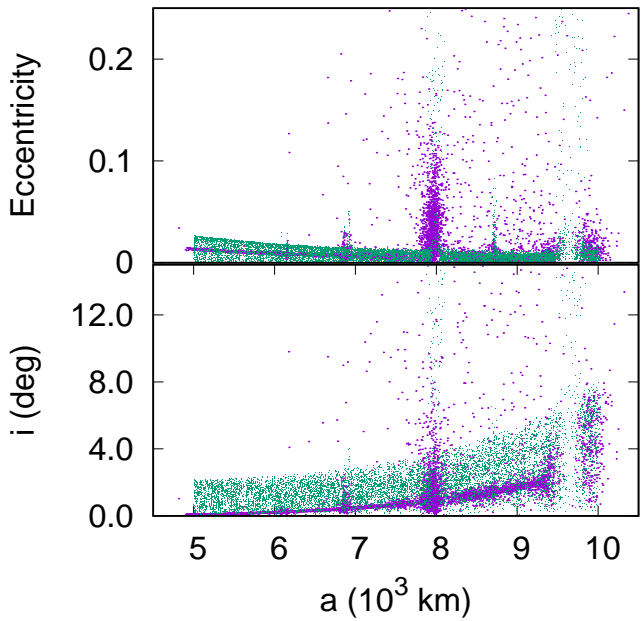}
\vskip -0.8 truecm
\caption{Orbital distribution of the ring particles after 250000 days from the beginning of the simulation in the case without collisions (green circles) and with collisions (magenta circles). The observed spikes in both $e$ and $i$ are due to mean motion resonances with the satellite.}
\label{fig:J2_KOZAI}
\end{figure}

When I included the effects of the J2 term in the potential of the central body, assuming a 2:1 ratio between the  $r_c$ and $r_a=r_b$ principal axes, the evolution is totally different. The ring maintains its coherence, even in the absence of collisions, and both the eccentricities and inclinations of the ring particles remain low, except at the locations of mean motion resonances with the satellite.
The protection against the Kozai--Lidov perturbations is due to the fast  precession of the pericenter argument induced by the J2 term.  The ring evolution is shown in Fig. \ref{fig:J2_KOZAI}, where I compare the cases with and without collisions. In both scenarios, the ring particles, after 250000 days, are still on 
low--eccentricity and low--inclination orbits. At the resonance locations with the satellite (3:1 at about 9600 km, 4:1 at about 8000 km, etc.), the particles are excited by the resonant perturbations and spikes in both orbital elements are observed.  

The critical distance within which the J2
disturbing potential term wins over the Kozai--Lidov mechanism 
can be estimated analytically  by comparing the precession rates induced by the two perturbations. The equation for the variation of the pericenter argument
is \citep{bertotti2003}:

\begin{equation}
\frac{d\omega}{dt} = -\frac{3}{4} n J_2 \left(\frac{R_c}{a}\right)^2 \frac{1}{(1-e^2)^2}(1 - 5\text{cos}^2 i )
.\end{equation}
\label{argument_pericenter}

In these equations, $n$ is the satellite mean motion, $R_c$ is the mean radius of the central body, and $a$, $e$, and $i$ are the semi--major axis, eccentricity, and   inclination of the ring particle, respectively. \\
The timescale associated with the precession of the pericenter argument can be estimated as $t_{J2} \sim \frac{2\pi}{d\omega/dt}$ and it is

\begin{equation}
 t_{J2} = \frac{8}{3}\pi \frac{1}{n J_2} \left(\frac{a}{R_c} \right)^2 {(1-e^2)^2} \frac{1}{5\text{cos}^2 i - 1}
.\end{equation}
\label{timescale_irregularity_inclination}

        If the orbital inclination is small, $cos(i) \sim 1$, the above equation can be approximated
as:
\begin{equation}
 t_{J2}|_{i=0} = \frac{2}{3}\pi \frac{1}{n J_2} \left( \frac{a}{R_c} \right)^2 {(1-e^2)^2}
.\end{equation}
\label{timescale_irregularity}

For the Kozai--Lidov perturbations, the timescale associated with the precession of the argument of pericenter, in the quadrupole approximation (small eccentricity of the outer perturber), is given by \citep{anto2015,naoz2016}
\begin{equation}
  t_\text{KL}= \frac{16}{15} \frac{M + m_s}{m_s} \frac{n}{n^2_s} (1-e_s^2)^{\frac{3}{2}}
,\end{equation}
\label{timescale_kozai}

where $m_s$ is the satellite mass,  $n_s$ is its mean motion, and $e_s$ is its eccentricity. 
By comparing $t_{KL}$ with $t_{J2}|_{i=0}$, a critical semi--major axis for which the two timescales are equal can be derived: 

\begin{equation}
 {{a}_c} = \left (\frac{8}{5\pi} R_c^2 J_2 \frac{(M + m_s)}{m_s} a_s^3 \frac{(1-e_s^2)^{\frac{3}{2}}}{(1-e^2)^2} \right)^{1/5} 
,\end{equation}

with $a_s$ being the semi--major axis of the satellite orbit. 

Within ${{a}_c}$, the J2 term will  prevail in determining the precession of the argument of the ring particle pericenter, and the Kozai--Lidov oscillations will be suppressed.  Beyond ${{a}_c}$, the Kozai--Lidov perturbations dominate over the J2 effects and the wide changes in eccentricity and inclination will be restored,  causing a progressive disruption of a putative ring. 

\begin{figure}[h!]

\includegraphics[width=0.5\columnwidth,angle=-90]{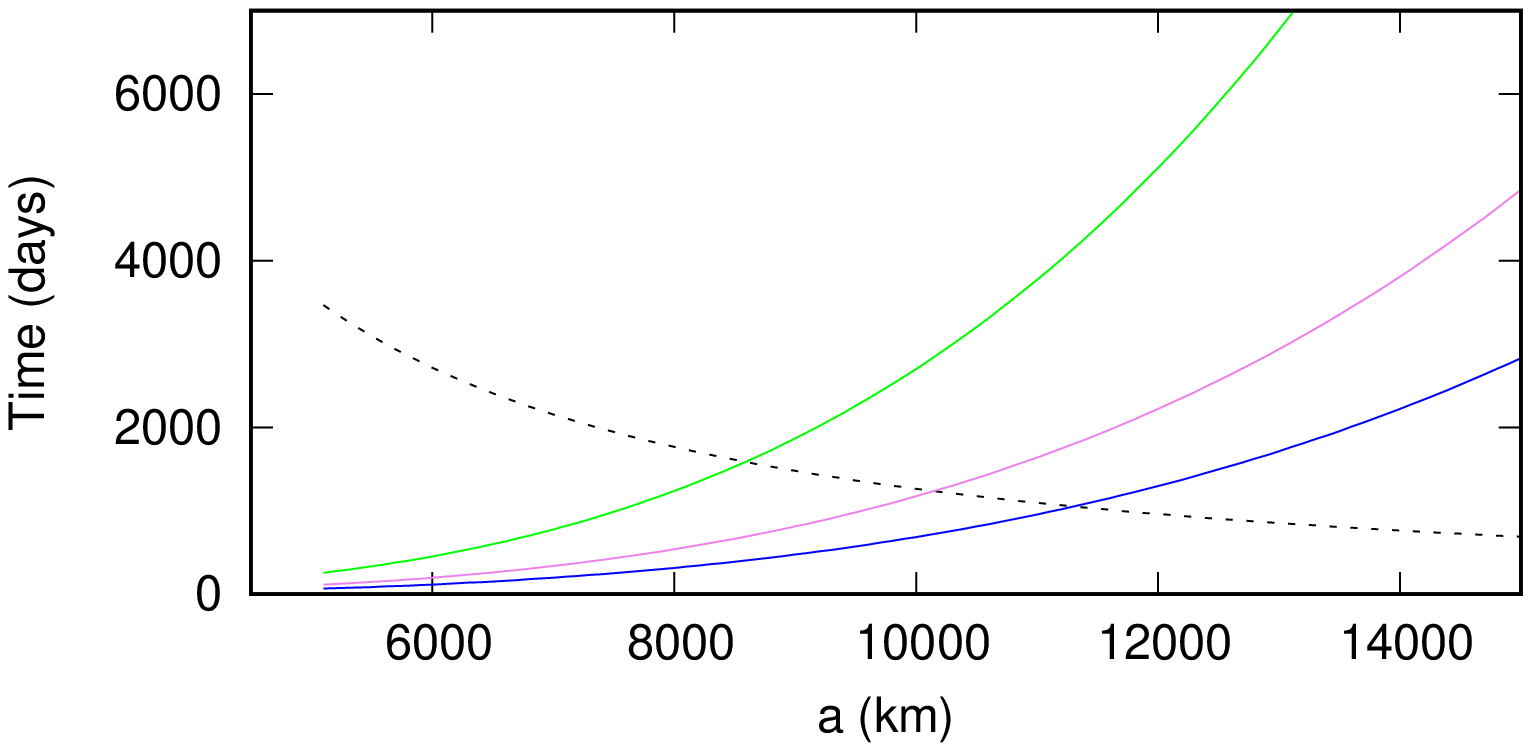}
\includegraphics[width=0.5\columnwidth,angle=-90]{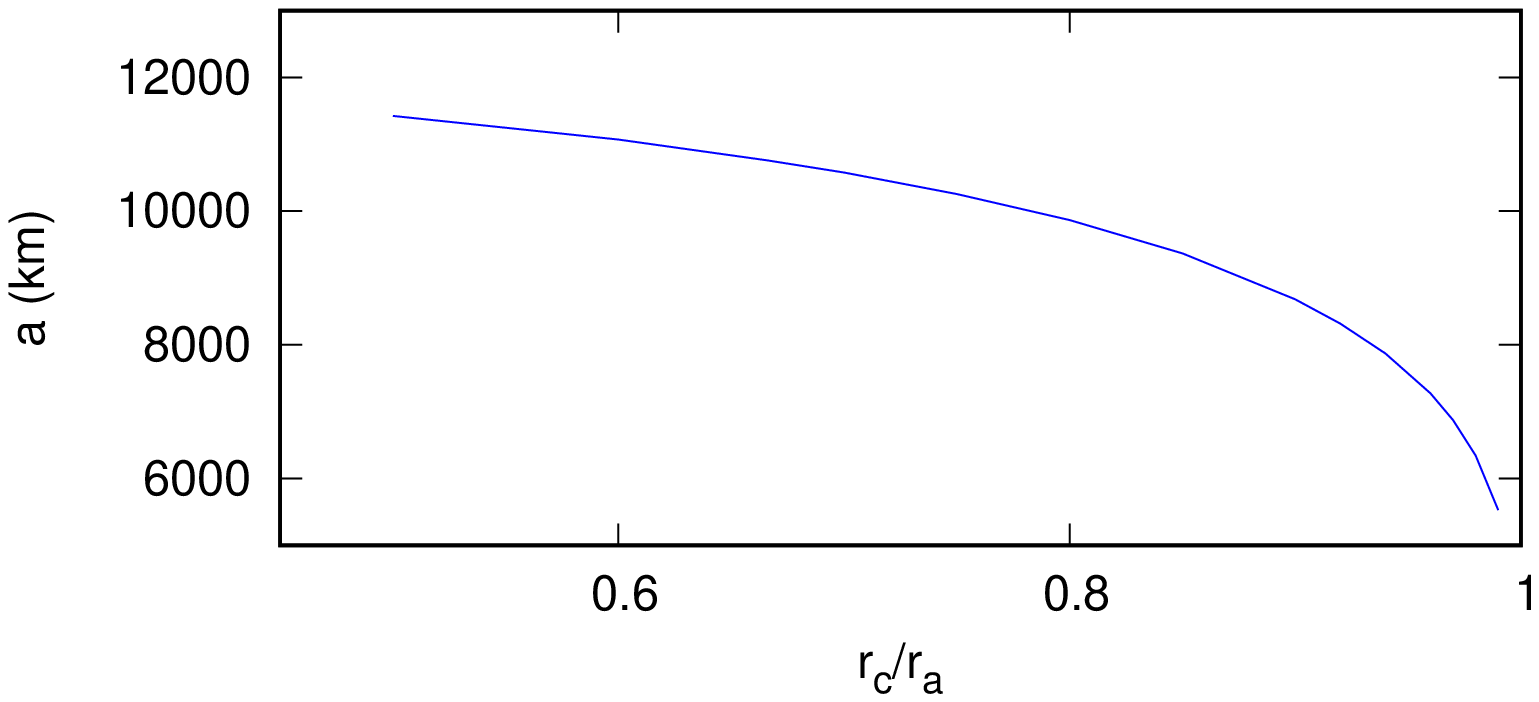}
\caption{Top panel: $t_{KL}$  shown as a function of the semi--major axis (dashed black line) and $t_{J2}|_{i=0} $  plotted for different values of the ratio between the principal axes of the  ellipsoid $r_c/r_a$. The blue line is for $r_c/r_a = 1/2$, the magenta line is for $r_c/r_a = 3/4$, and the green line is for  $r_c/r_a = 9/10$. 
Bottom panel:  critical semi--major axis drawn as a function of the $r_c/r_a$ ratio. }
\label{fig:J2_KOZAI_ANAL}
\end{figure}

In the top panel of Fig. \ref{fig:J2_KOZAI_ANAL}, the two timescales $t_{KL}$ and $t_{J2}|_{i=0}$ are compared for different values of the ratio between $r_c$ and $r_a$, showing that a larger flattening of the central body shifts the critical semi--major axis, within which  the ring is stable against the Kozai--Lidov perturbations, outward. In the bottom panel, the value of ${{a}_c}$ is computed for decreasing values of the $r_c/r_a$, showing the range in which the J2 perturbations can protect an equatorial ring.  
These calculations show that a significant oblateness of the central body is crucial to ensuring the survival of an equatorial ring around the body in the presence of a highly inclined satellite. 

\section{Secular perturbations of an inclined satellite and final ring inclination}

\begin{figure}[h!]
\hskip -1.9 truecm
\includegraphics[width=1.2\columnwidth,angle=-90]{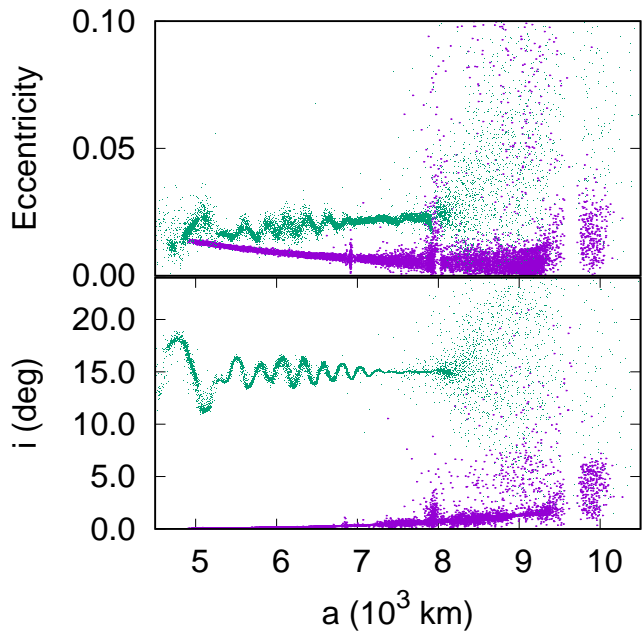}
\vskip -0.6 truecm
\caption{Orbital distribution of the ring particles after 150000 days in the J2 term  (magenta circles) and spherical cases (green circles). As in Fig. \ref{fig:J2_KOZAI}, the observed spikes in both $e$ and $i$ are due to mean motion resonances with the satellite.}
\label{fig:J2_SEC}
\end{figure}

\begin{figure}[h!]
\hskip -0.2 truecm
\includegraphics[width=0.5\columnwidth,angle=-90]{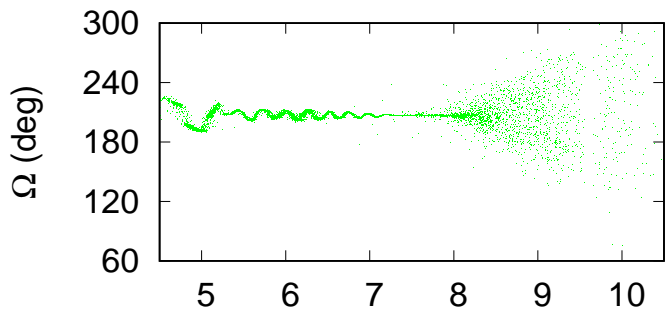}
\caption{Distribution of the node longitudes of the ring particles after 150000 days in the case without the J2 term. The alignment, at least up to 8000 km, ensures the compactness of the ring. }
\label{fig:NODE_SEC}
\end{figure}

If the inclination of the perturbing satellite is mild (lower than $\sim 39^o$), the evolution of the inclination and eccentricity is less dramatic. I ran a model with the inclination of the perturber set to $15^o$,  switching the contribution from the J2 term on and off; the results
are shown in Fig. \ref{fig:J2_SEC}. In the case of a spherical central body, the combined effect of collisions and secular perturbations
by the satellite forces the ring to relax to the same orbital plane as the satellite. As illustrated in   Fig.\ref{fig:J2_SEC}, the
inclination of the ring particles has small oscillations around the inclination of the satellite, which may be further damped at later times, while the nodal longitudes are aligned; this dynamical configuration maintains the coherence of the ring (Fig. \ref{fig:NODE_SEC}).  Collisions act as a viscous force that tends to align the ring to the plane of the perturber. The ring appears slightly warped close to the central body
(see Fig. \ref{fig:J2_SEC})  while it is well aligned in the central part. The warping is probably a temporary feature and may
disappear at later times. 

 In the J2 term case, the eccentricities and inclinations are on average always small, and the ring is coherent and does not depart from the equatorial plane of the central body. The outer regions of the disk, approximately 8000 km beyond, show some scattering around the middle plane where the protective effect of the J2 term weakens. 
An analytical estimate of the limiting semi--major axis, within which the J2 holds the ring in the equatorial plane, is more complex in this case. The second order Lagrange--Laplace secular theory is a good approximation for low values of eccentricity and inclination. In this particular case, the inclination is set to $15^o$ and the theory may not be very accurate. However, this secular theory can give an idea of the timescales involved in the dynamical evolution of the ring. According to \cite{murray-dermottSS}, the proper oscillation timescale for the pericenter longitude is given by $T_{sec} = 2 \pi / A$, where A reads:

\begin{equation}
A = n \frac{1} {4} \frac {m_{1}} {M_c} \left (\frac {a} {a_s} \right )^2 b_{3/2}^{(1)}
.\end{equation}

\begin{figure}[h!]
\includegraphics[width=0.5\columnwidth,angle=-90]{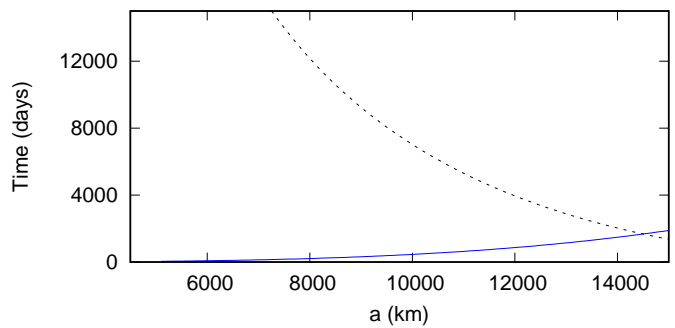}
\caption{Comparison of the timescale for the secular precession of the pericenter longitude (dashed black line) and the timescale due to the J2 term (blue continuous line).}
\label{fig:J2_SEC_ANAL}
\end{figure}

This frequency has a complex dependence on the semi--major axis of the ring particle due to the Laplace coefficient
$b_{3/2}^{(1)}$. For this reason, numerical computations lead to an easier evaluation of  the value of semi--major axis ${{a}_c}$, for which the two time derivatives of the pericenter longitude, the one due to the J2 perturbations and the secular one, match.  In  Fig. \ref{fig:J2_SEC_ANAL}, the two timescales associated with the frequencies are compared. The critical semi--major axis is located at approximately 
15000 km and is in good agreement with the numerical modeling (Fig. \ref{fig:J2_SEC}). 

\section{Disruption of a ring misaligned with respect to the equatorial plane of the central body}

\begin{figure}[h!]

\includegraphics[width=0.7\columnwidth,angle=-90]{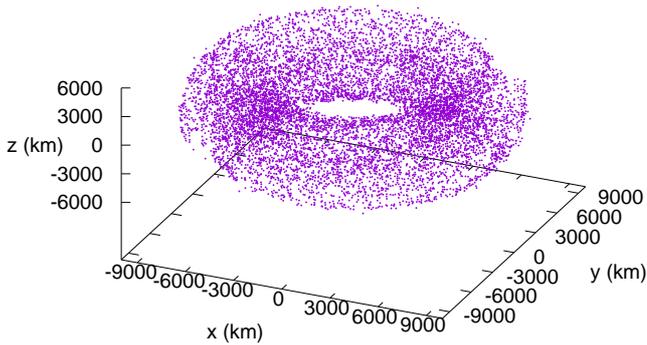}

\caption{Three-dimensional distribution of the ring particles after 2000 days. The nodal precession forced by J2 transforms the ring into a torus.}
\label{fig:3D}
\end{figure}

If the ring is formed on a plane that is tilted with respect to the equatorial plane of an oblate central body, the J2 term quickly disrupts the ring. The forced circulation of the nodal longitudes de--phases the inclined orbits of the ring particles, and the initial ring is transformed into a  torus. The fattening of the ring depends on its initial inclination  with respect to the equatorial plane. In Fig. \ref{fig:3D}, the three--dimensional distribution of the ring particles is shown. They quickly disperse on a timescale on the order of a few $10^3$ yr and do not form a flat coherent structure anymore. 
In this case, the J2 perturbation is not a protective factor
but rather a disruptive influence.
As a consequence, potential rings around an oblate body must be sought in the equatorial plane.   

\section{The case of Haumea: Interplay among its complex shape, fast rotation, and two satellites} 

Haumea, a dwarf planet orbiting beyond Neptune, has a very elongated shape and a fast rotation around the principal  axis of inertia ($\sim 3.92$ h, \cite{lello2010}). It is modeled as a triaxial ellipsoid, and its  semi--axes  are
$r_a =1161 \pm 30$ km,  $r_b = 852 \pm 4$ km, and 
$r_c = 513 \pm 16$ km \citep{ortiz2017} with a ratio between $r_c$ and $r_a$ of about 1:2 (same as the test case adopted in the previous sections). Haumea is orbited by two satellites, Namaka and Hi'iaka, and by a ring coplanar with the equator of Haumea with a radius of $\sim 2287$ km and a width of $\sim 70$ km \citep{ortiz2017}.  According to the previous results, the coplanarity of the ring with the equator of Haumea is expected since the body is a strongly oblate body (see Sect. 5). 

The more massive satellite Hi'iaka (its mass is approximately 1/223 times the mass of Haumea) has a semi--major axis of 49880 km \citep{ragozzine2009}; it is almost on a circular orbit ($e \sim 0.05$) and, according to \cite{ortiz2017}, has an orbital plane coplanar with the equator of Haumea. One therefore expects that its perturbations on the inclination of the ring are negligible. Namaka has a mass about ten times smaller than that of Hi'iaka, but it is closer to Haumea 
since its semi--major axis is 25657 km, about half that of Hi'iaka.
It is on an eccentric ($e \sim 0.25$) and inclined ($i \sim 13^o$) orbit with respect to the equatorial plane of Haumea, and it may cause long-term perturbations on the inclination of the ring particles, potentially dispersing them. 

\begin{figure}[h!]

\includegraphics[width=0.47\columnwidth,angle=-90]{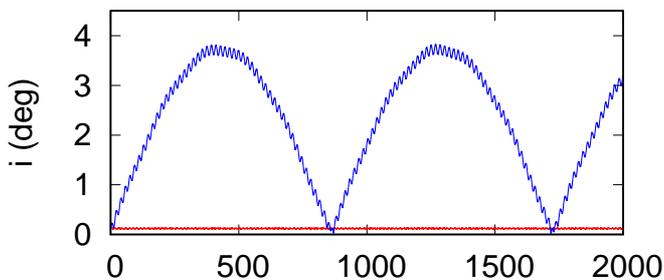}

\caption{Time evolution of the inclination of a ring particle for a spherical central body (blue dashed line) and for a rotating triaxial ellipsoid (red line), like the real Haumea.}
\label{fig:haumea1}
\end{figure}

In Fig. \ref{fig:haumea1}, the evolution of a single particle initially set in the middle of the ring is shown in two different configurations. The blue dashed line shows the inclination evolution of the particle for a spherical central body ($r_a=r_b=r_c$) while the red line shows the evolution of the same particle when the real values of $r_a, r_b,$ and $r_c$ and the rotation of the body are included. In the first case, the inclination of the particle regularly oscillates between $1^o$ and $4^o$  due to the secular perturbations of Namaka. As expected from Sect. 4, when the J2 term is included and I consider the full model for Haumea, the inclination is almost constant with very small oscillations around the initial value.  Even in the case of Haumea's ring, the J2 term helps in keeping the ring coherent in the equatorial plane of the body.  

\begin{figure}[h!]
\hskip -1.5 truecm
\includegraphics[width=1.2\columnwidth,angle=-90]{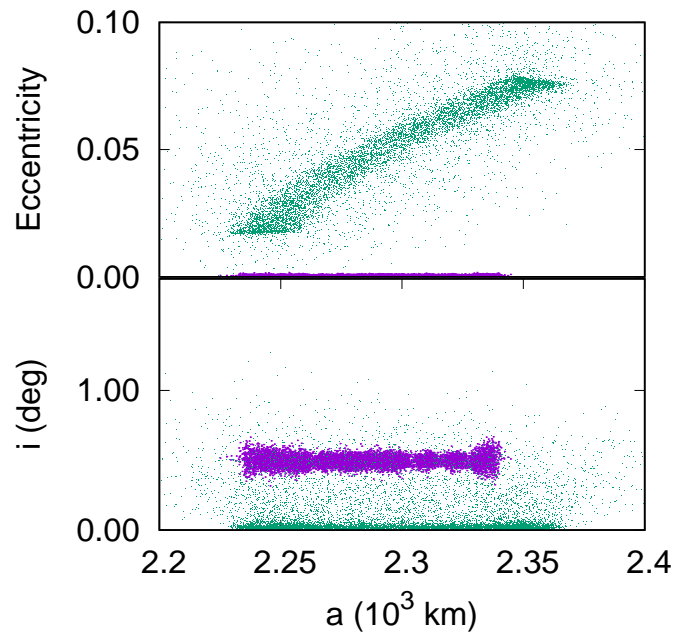}

\caption{Orbital distribution of the ring particles after $10^4$ days. In the top panel, the eccentricity vs. semi--major axis is shown in two different models. The magenta dots illustrate the case where $B=A \not= C$ (only the J2 term) while the green ones show the outcome of the full model. The inclination vs. semi--major axis is shown in the top panel.  }
\label{fig:haumea2}
\end{figure}

To explore the evolution of the ring, also including the effects of mutual collisions, I generated a population of 10000 particles with initial semi--major axes between 2235 and 2340 km, eccentricity lower than $10^{-4 }$, and inclination lower than $0.5^o$ (both randomly selected). An inflated diameter of  $3$ km was adopted in order to attain significant collisional activity. 
In Fig. \ref{fig:haumea2}, the eccentricity and inclination distribution is compared after $10^4$ days of evolution in the case where only the J2 term is included ($r_b-r_a=0$) and in the real case where $r_b-r_a \not = 0$. When $r_b=r_a$, the eccentricity of all particles in the ring is very low and the magenta dots, which represent them in  Fig. \ref{fig:haumea2}, are barely visible above zero. The inclination does not change significantly from the initial value and, as shown in Fig. \ref{fig:haumea1}, it is not expected to grow since the J2 perturbations suppress the secular perturbations of Namaka. When the perturbations due to  $r_b \not = r_a$ are included (full model), the potential of the central body excites the eccentricities of the ring particles to higher values, as shown in the top panel of Fig. \ref{fig:haumea2}, despite the damping effect of collisions. This is potentially related to the 3:1 resonance between the rotation period of Haumea and the orbital period of the ring particles  \citep{winter2019}. On the other hand, the inclination is on average smaller than the case with J2--only, as shown in the bottom panel of Fig. \ref{fig:haumea2}. 

\begin{figure}[h!]

\includegraphics[width=0.5\columnwidth,angle=-90]{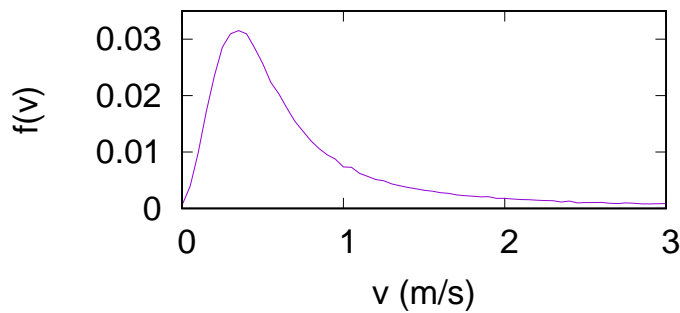}

\caption{Distribution of the mutual impact velocities between the ring particles of Haumea when a collisional steady state is reached.  }
\label{fig:haumea3}
\end{figure}

The high values of eccentricity induced by the $r_a \neq r_b$ term of the potential lead to an increase in the mutual velocities between the ring particles. This may speed up the erosion of the ring if they are too high and may possibly cause fragmentation of the impactors most of the time. In Fig. \ref{fig:haumea3}, I show the normalized distribution of the impact velocities when the steady state is reached (Fig. \ref{fig:haumea2}). The large majority of collisions occur at a speed lower than 1 m/s, which, according to \cite{syed2017}, is below the erosion limit for cm--size particles. Therefore, the increase in the impact velocity due to the high eccentricity, possibly related to the 3:1 resonance, is not enough to cause a quick erosion of Haumea's ring. 

\section{Conclusions}

The dynamical evolution of a ring around a minor body or dwarf planet that also possesses one or more satellites is complex. Different dynamical mechanisms come into play and may  lead to the quick disruption of the ring. If the satellite is inclined with respect to the ring plane, the oscillations in inclination, in particular in the Kozai--Lidov regime, can fully disperse the ring on a short timescale. 
In addition, if the central body is significantly oblate, the fast pericenter precession induced by the J2 term 
quickly disperses the putative rings
that are significantly inclined with respect to the equatorial plane. 


To determine under which conditions a ring can survive, I exploited 
a numerical model that  includes the oblateness of the central body, the mutual collisions between the ring particles, and the gravitational perturbations of one or more satellites.  Analytical calculations
have also been performed to evaluate the relative strength and contribution 
of the different perturbing mechanisms. 

I find that the J2  term is double--faced. It is disruptive for inclined rings, but it 
protects equatorial rings from the perturbations of an inclined satellite. This occurs for the strong Kozai--Lidov oscillations and also for the milder secular perturbations that occur when the inclination is lower than $39^o$.
The J2 term suppresses the inclination perturbations of the satellite  within a critical semi--major axis, which depends on the oblateness of the central body, by forcing a fast pericenter precession that dominates over that induced by the secular terms of the satellite. This is a general result that can be applied to any system, and it allows us to predict that rings formed around oblate bodies must be preferentially searched for in the equatorial plane. Rings that are born inclined with respect to this plane can survive only if the central body has a small J2. The dynamical structure of the central body and satellite determines which kind of rings can survive after their formation. 

The  Haumea case was also investigated with numerical simulations, and I find that its ring is forced by the strong J2 perturbations of the oblate triaxial ellipsoid to lie in the equatorial plane. In addition, the $r_b \neq r_a$ gravitational perturbations excite large eccentricities among the ring particles that are not damped by collisions. However, the relative impact velocities are lower than the expected fragmentation velocity and the survival of the ring is guaranteed.  

\begin{acknowledgements}
I would like to thank the referee, Anthony Dobrovolskis, for his comments
and suggestions and Mariasole Maglione for her help with the 
analytical calculations. 
\end{acknowledgements}

\bibliographystyle{aa}
\bibliography{biblio}

\begin{thebibliography}{22}
\expandafter\ifx\csname natexlab\endcsname\relax\def\natexlab#1{#1}\fi

\bibitem[{{Antognini}(2015)}]{anto2015}
{Antognini}, J.~M.~O. 2015, \mnras, 452, 3610

\bibitem[{{Arakawa} {et~al.}(2019){Arakawa}, {Hyodo}, \& {Genda}}]{arakawa2019}
{Arakawa}, S., {Hyodo}, R., \& {Genda}, H. 2019, Nature Astronomy, 3, 802

\bibitem[{{Bertotti} {et~al.}(2003){Bertotti}, {Farinella}, \&
  {Vokrouhlick}}]{bertotti2003}
{Bertotti}, B., {Farinella}, P., \& {Vokrouhlick}, D. 2003, {Physics of the
  Solar System - Dynamics and Evolution, Space Physics, and Spacetime
  Structure.}, Vol. 293

\bibitem[{{Braga-Ribas} {et~al.}(2014){Braga-Ribas}, {Sicardy}, {Ortiz},
  {Snodgrass}, {Roques}, {Vieira-Martins}, {Camargo}, {Assafin}, {Duffard},
  {Jehin}, {Pollock}, {Leiva}, {Emilio}, {Machado}, {Colazo}, {Lellouch},
  {Skottfelt}, {Gillon}, {Ligier}, {Maquet}, {Benedetti-Rossi}, {Gomes},
  {Kervella}, {Monteiro}, {Sfair}, {El Moutamid}, {Tancredi}, {Spagnotto},
  {Maury}, {Morales}, {Gil-Hutton}, {Roland}, {Ceretta}, {Gu}, {Wang},
  {Harps{\o}e}, {Rabus}, {Manfroid}, {Opitom}, {Vanzi}, {Mehret}, {Lorenzini},
  {Schneiter}, {Melia}, {Lecacheux}, {Colas}, {Vachier}, {Widemann},
  {Almenares}, {Sand ness}, {Char}, {Perez}, {Lemos}, {Martinez},
  {J{\o}rgensen}, {Dominik}, {Roig}, {Reichart}, {Lacluyze}, {Haislip},
  {Ivarsen}, {Moore}, {Frank}, \& {Lambas}}]{ribas2014}
{Braga-Ribas}, F., {Sicardy}, B., {Ortiz}, J.~L., {et~al.} 2014, \nat, 508, 72

\bibitem[{{Brahic}(1976)}]{brahic1976}
{Brahic}, A. 1976, Journal of Computational Physics, 22, 171

\bibitem[{{Brahic}(1977)}]{brahic1977}
{Brahic}, A. 1977, \aap, 54, 895

\bibitem[{{Bukhari Syed} {et~al.}(2017){Bukhari Syed}, {Blum}, {Wahlberg
  Jansson}, \& {Johansen}}]{syed2017}
{Bukhari Syed}, M., {Blum}, J., {Wahlberg Jansson}, K., \& {Johansen}, A. 2017,
  \apj, 834, 145

\bibitem[{{Charnoz} {et~al.}(2001){Charnoz}, {Th{\'e}bault}, \&
  {Brahic}}]{charnoz2001}
{Charnoz}, S., {Th{\'e}bault}, P., \& {Brahic}, A. 2001, \aap, 373, 683

\bibitem[{{Everhart}(1985)}]{radau1985}
{Everhart}, E. 1985, Astrophysics and Space Science Library, Vol. 115, {An
  efficient integrator that uses Gauss-Radau spacings}, ed. A.~{Carusi} \&
  G.~B. {Valsecchi}, 185

\bibitem[{{Hyodo} {et~al.}(2016){Hyodo}, {Charnoz}, {Genda}, \&
  {Ohtsuki}}]{ryuki2016}
{Hyodo}, R., {Charnoz}, S., {Genda}, H., \& {Ohtsuki}, K. 2016, \apjl, 828, L8

\bibitem[{{Lellouch} {et~al.}(2010){Lellouch}, {Kiss}, {Santos-Sanz},
  {M{\"u}ller}, {Fornasier}, {Groussin}, {Lacerda}, {Ortiz}, {Thirouin},
  {Delsanti}, {Duffard}, {Harris}, {Henry}, {Lim}, {Moreno}, {Mommert},
  {Mueller}, {Protopapa}, {Stansberry}, {Trilling}, {Vilenius}, {Barucci},
  {Crovisier}, {Doressoundiram}, {Dotto}, {Guti{\'e}rrez}, {Hainaut},
  {Hartogh}, {Hestroffer}, {Horner}, {Jorda}, {Kidger}, {Lara}, {Rengel},
  {Swinyard}, \& {Thomas}}]{lello2010}
{Lellouch}, E., {Kiss}, C., {Santos-Sanz}, P., {et~al.} 2010, \aap, 518, L147

\bibitem[{{Murray} \& {Dermott}(1999)}]{murray-dermottSS}
{Murray}, C.~D. \& {Dermott}, S.~F. 1999, {Solar System Dynamics}

\bibitem[{{Naoz}(2016)}]{naoz2016}
{Naoz}, S. 2016, \araa, 54, 441

\bibitem[{{Ortiz} {et~al.}(2015){Ortiz}, {Duffard}, {Pinilla-Alonso},
  {Alvarez-Cand al}, {Santos-Sanz}, {Morales}, {Fernand ez-Valenzuela},
  {Licandro}, {Campo-Bagatin}, \& {Thirouin}}]{ortiz2015}
{Ortiz}, J.~L., {Duffard}, R., {Pinilla-Alonso}, N., {et~al.} 2015, in European
  Planetary Science Congress, EPSC2015--230

\bibitem[{{Ortiz} {et~al.}(2017){Ortiz}, {Santos-Sanz}, {Sicardy},
  {Benedetti-Rossi}, {B{\'e}rard}, {Morales}, {Duffard}, {Braga-Ribas}, {Hopp},
  {Ries}, {Nascimbeni}, {Marzari}, {Granata}, {P{\'a}l}, {Kiss}, {Pribulla},
  {Kom{\v{z}}{\'\i}k}, {Hornoch}, {Pravec}, {Bacci}, {Maestripieri}, {Nerli},
  {Mazzei}, {Bachini}, {Martinelli}, {Succi}, {Ciabattari}, {Mikuz},
  {Carbognani}, {Gaehrken}, {Mottola}, {Hellmich}, {Rommel},
  {Fern{\'a}ndez-Valenzuela}, {Campo Bagatin}, {Cikota}, {Cikota}, {Lecacheux},
  {Vieira-Martins}, {Camargo}, {Assafin}, {Colas}, {Behrend}, {Desmars},
  {Meza}, {Alvarez-Candal}, {Beisker}, {Gomes-Junior}, {Morgado}, {Roques},
  {Vachier}, {Berthier}, {Mueller}, {Madiedo}, {Unsalan}, {Sonbas}, {Karaman},
  {Erece}, {Koseoglu}, {Ozisik}, {Kalkan}, {Guney}, {Niaei}, {Satir},
  {Yesilyaprak}, {Puskullu}, {Kabas}, {Demircan}, {Alikakos}, {Charmandaris},
  {Leto}, {Ohlert}, {Christille}, {Szak{\'a}ts}, {Tak{\'a}csn{\'e} Farkas},
  {Varga-Vereb{\'e}lyi}, {Marton}, {Marciniak}, {Bartczak}, {Santana-Ros},
  {Butkiewicz-B{\k{a}}k}, {Dudzi{\'n}ski}, {Al{\'\i}-Lagoa}, {Gazeas},
  {Tzouganatos}, {Paschalis}, {Tsamis}, {S{\'a}nchez-Lavega},
  {P{\'e}rez-Hoyos}, {Hueso}, {Guirado}, {Peris}, \&
  {Iglesias-Marzoa}}]{ortiz2017}
{Ortiz}, J.~L., {Santos-Sanz}, P., {Sicardy}, B., {et~al.} 2017, \nat, 550, 219

\bibitem[{{Pan} \& {Wu}(2016)}]{pan2016}
{Pan}, M. \& {Wu}, Y. 2016, \apj, 821, 18

\bibitem[{{Ragozzine} \& {Brown}(2009)}]{ragozzine2009}
{Ragozzine}, D. \& {Brown}, M.~E. 2009, \aj, 137, 4766

\bibitem[{{Ruprecht} {et~al.}(2015){Ruprecht}, {Bosh}, {Person}, {Bianco},
  {Fulton}, {Gulbis}, {Bus}, \& {Zangari}}]{ruprecht2015}
{Ruprecht}, J.~D., {Bosh}, A.~S., {Person}, M.~J., {et~al.} 2015, \icarus, 252,
  271

\bibitem[{{Salmon} {et~al.}(2010){Salmon}, {Charnoz}, {Crida}, \&
  {Brahic}}]{salmon}
{Salmon}, J., {Charnoz}, S., {Crida}, A., \& {Brahic}, A. 2010, \icarus, 209,
  771

\bibitem[{{Th{\'e}bault} \& {Brahic}(1998)}]{theb1998}
{Th{\'e}bault}, P. \& {Brahic}, A. 1998, \planss, 47, 233

\bibitem[{Turcotte {et~al.}(2002)Turcotte, Schubert, \& Schubert}]{turcotte}
Turcotte, D., Schubert, G., \& Schubert, J. 2002, Geodynamics (Cambridge
  University Press)

\bibitem[{{Winter} {et~al.}(2019){Winter}, {Borderes-Motta}, \&
  {Ribeiro}}]{winter2019}
{Winter}, O.~C., {Borderes-Motta}, G., \& {Ribeiro}, T. 2019, \mnras, 484, 3765

\end{thebibliography}

\end{document}